\documentclass[iop, apj, twocolappendix, numberedappendix, appendixfloats]{emulateapj}
\usepackage{threeparttable}
\usepackage{multirow}
\usepackage{times}
\usepackage{enumitem}
\usepackage{url} 
\usepackage{amsmath,bm}

\shorttitle{}
\shortauthors{Y. Huang et al.}

\begin{document}

\title{Discovery of A candidate Hypervelocity star originated from the Sagittarius Dwarf Spheroidal galaxy}

\author{Yang Huang\altaffilmark{1,8}}
\author{Qingzheng Li\altaffilmark{2,3,8}}
\author{Huawei Zhang\altaffilmark{4,5}}
\author{Xinyi Li\altaffilmark{1}}
\author{Weixiang Sun\altaffilmark{1}}
\author{Jiang Chang\altaffilmark{6,7}}
\author{Xiaobo Dong\altaffilmark{2}}
\author{Xiaowei Liu\altaffilmark{1,8}}

\altaffiltext{1}{South-Western Institute for Astronomy Research, Yunnan University, Kunming 650500, P.\,R.\,China; {\it yanghuang@ynu.edu.cn;  x.liu@ynu.edu.cn}}
\altaffiltext{2}{Yunnan Observatories, Chinese Academy of Sciences, Kunming 650011, P.\,R.\,China; {\it liqingzheng@ynao.ac.cn}}
\altaffiltext{3}{University of Chinese Academy of Sciences, Beijing 100049, P.\,R.\,China}
\altaffiltext{4}{Department of Astronomy, School of Physics, Peking University, Beijing 100871, P.\,R.\,China}
\altaffiltext{5}{Kavli Institute for Astronomy and Astrophysics, Peking University, Beijing 100871, P.\,R.\,China}
\altaffiltext{6}{Key Lab of Optical Astronomy, National Astronomical Observatories, Chinese Academy of Sciences, Beijing 100012, P.\,R.\,China}
\altaffiltext{7}{Purple Mountain Observatory, Chinese Academy of Sciences, Nanjing 210034, P.\,R.\,China}
\altaffiltext{8}{Corresponding authors}

\begin{abstract}
In this letter, we report the discovery of an intriguing HVS (J1443+1453) candidate that is probably from the Sagittarius Dwarf Spheroidal galaxy (Sgr dSph).
The star is an old and very metal-poor low-mass main-sequence turn-off star (age\,$\sim14.0$\,Gyr and [Fe/H]\,$= -2.23$\,dex) and has a total velocity of $559.01^{+135.07}_{-87.40}$\,km\,s$^{-1}$ in the Galactic rest-frame and a heliocentric distance of $2.90^{+0.72}_{-0.48}$\,kpc.
The velocity of J1443+1453 is larger than the escape speed at its position, suggesting it a promising HVS candidate.
By reconstructing its trajectory in the Galactic potential, we find that the orbit of  J1443+1453 intersects closely with that of the Sgr dSph $37.8^{+4.6}_{-6.0}$\,Myr ago, when the latter has its latest pericentric passage through the Milky Way. 
The encounter occurs at a distance $2.42^{+1.80}_{-0.77}$\,kpc from the centre of Sgr dSph, smaller than the size of the Sgr dSph.
Chemical properties of this star are also consistent with those of one Sgr dSph associated globular cluster or of the Sgr stream member stars.
Our finding suggests that J1443+1453 is an HVS either tidally stripped from the Sgr dSph or ejected from the Sgr dSph by the gravitational slingshot effect, requiring a (central) massive/intermediate-mass black hole or a (central) massive primordial black hole in the Sgr dSph.
\end{abstract}

\section{Introduction}
The existence of escaping hypervelocity stars (HVSs) in our Milky Way (MW) was first proposed by Hills (1988), as a consequence of the dynamical interactions between a stellar binary and the central super massive black hole (SMBH).
The interactions can propel one member of the binary and accelerate it to a speed exceeding 1000\,km\,s$^{-1}$, high enough to escape the gravitational potential of our Galaxy.
The first HVS, a B-type star with an extreme radial velocity of 709\,km\,s$^{-1}$ in the Galactic rest frame,  was discovered serendipitously in a spectroscopic survey of faint blue horizontal-branch (BHB) star candidates in the Galactic halo (Brown et al. 2005).
The star, together with two dozen early-type HVSs discovered in the follow-up dedicated surveys (Brown et al. 2006, 2009, 2012, 2014; Zheng et al. 2014; Huang et al. 2017), provide strong support of the Hills mechanism.
Most recently, the currently fastest HVS was discovered serendipitously by the Southern Stellar Stream Spectroscopic Survey (S$^5$) with a total velocity of $1755 \pm 50$\,km\,s$^{-1}$ located at a heliocentric distance of $\sim 9$\,kpc (Koposov et al. 2020).
By integrating its backward trajectory, this star was found to point unambiguously to the Galactic Centre, providing a direct evidence to the Hills mechanism.

\begin{table}
\caption{The measured parameters of HVS candidate J1443+1453.}
\centering
\begin{tabular}{lccc}
\hline
Parameter&Value&Units\\
\hline
Right Ascension (J2000)&14:43:25.76&\\
Declination (J2000)&$+$14:53:36.3&\\
{\it Gaia} DR2 source\_id&1186023710910901760&\\
{\it Gaia} DR2 Proper motion $\mu_{\alpha}\cos\delta$&$-46.914 \pm 0.121$&mas\,yr$^{-1}$\\
{\it Gaia} DR2 Proper motion $\mu_{\delta}$&$-1.465 \pm 0.096$&mas\,yr$^{-1}$\\
{\it Gaia} DR2 Parallax&$0.348 \pm 0.065$&mas\\
{\it Gaia} DR2 $G$-band magnitude&$16.072 \pm 0.002$&mag\\
{\it Gaia} DR2 $G_{\rm BP} - G_{\rm RP}$&$0.627 \pm 0.009$&mag\\
Distance&$2.90^{+0.72}_{-0.48}$&kpc\\
SDSS $g$-band magnitude&$16.321 \pm 0.004$&mag\\
SDSS $r$-band magnitude&$16.066 \pm 0.004$&mag\\
Colour excess $E (B - V)_{\rm SFD}$&$0.016$&mag\\
Heliocentric radial velocity&$194.25 \pm 1.97$&km\,s$^{-1}$\\
Effective temperature $T_{\rm eff}$&$6405 \pm 130$&K\\
Surface gravity log\,$g$&$3.64 \pm 0.21$&dex\\
Metallicity \text{[Fe/H]}&$-2.23 \pm 0.11$&dex\\
$\alpha$-element to iron ratio \text{[$\alpha$/Fe]}&$0.14 \pm 0.10$&dex\\
Galactocentric distance $r_{\rm GC}$&$7.30^{+0.31}_{-0.34}$&kpc\\
Total velocity $V_{\rm GSR}$&$559.01^{+135.07}_{-87.40}$&km\,s$^{-1}$\\
Age&$14.0^{+0.5}_{-2.0}$&Gyr\\
Mass&$0.75^{+0.03}_{-0.01}$&$M_\odot$\\
\hline
\end{tabular}
\end{table}

 \begin{figure*}
\begin{center}
\includegraphics[scale=0.35,angle=0]{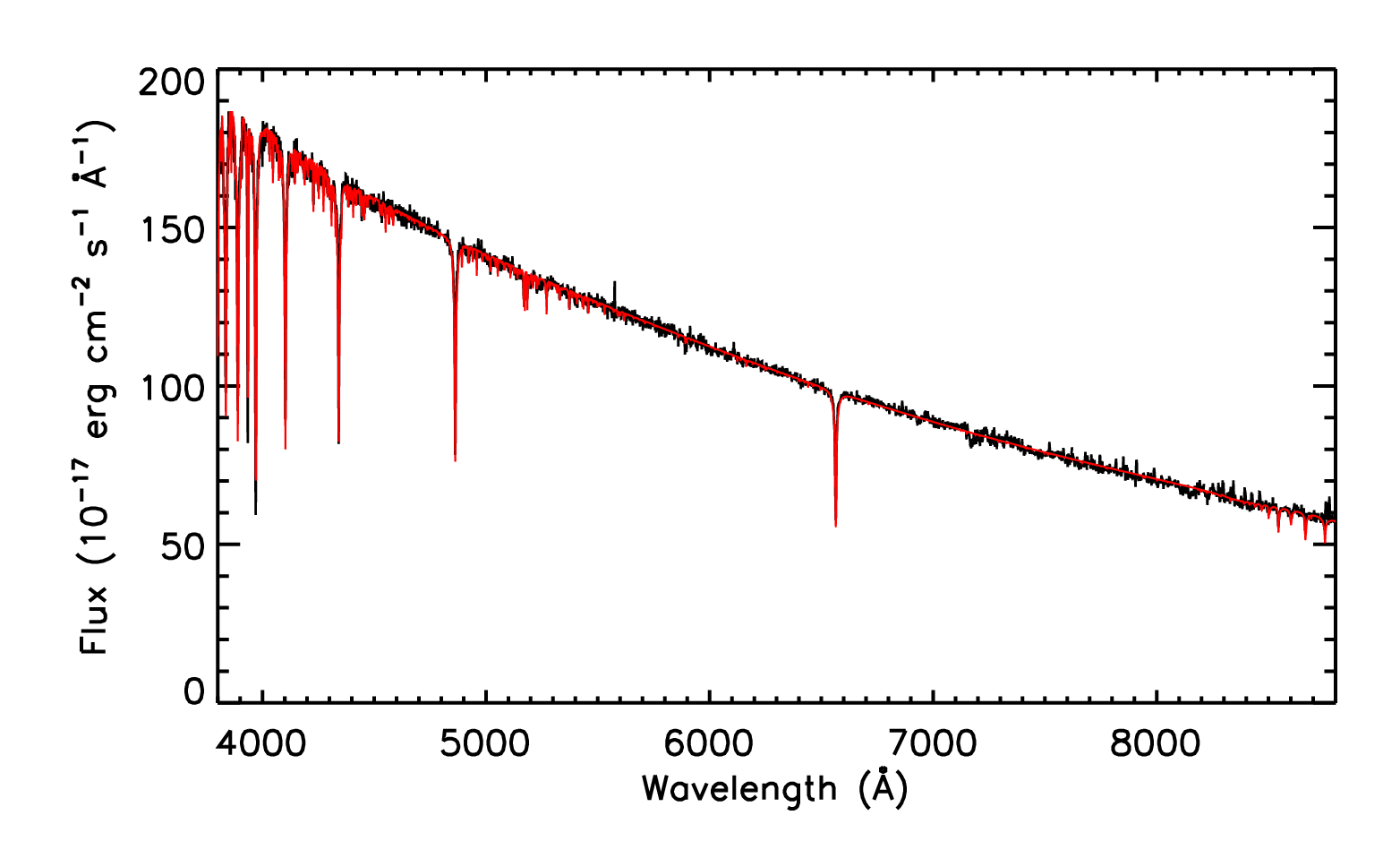}
\includegraphics[scale=0.35,angle=0]{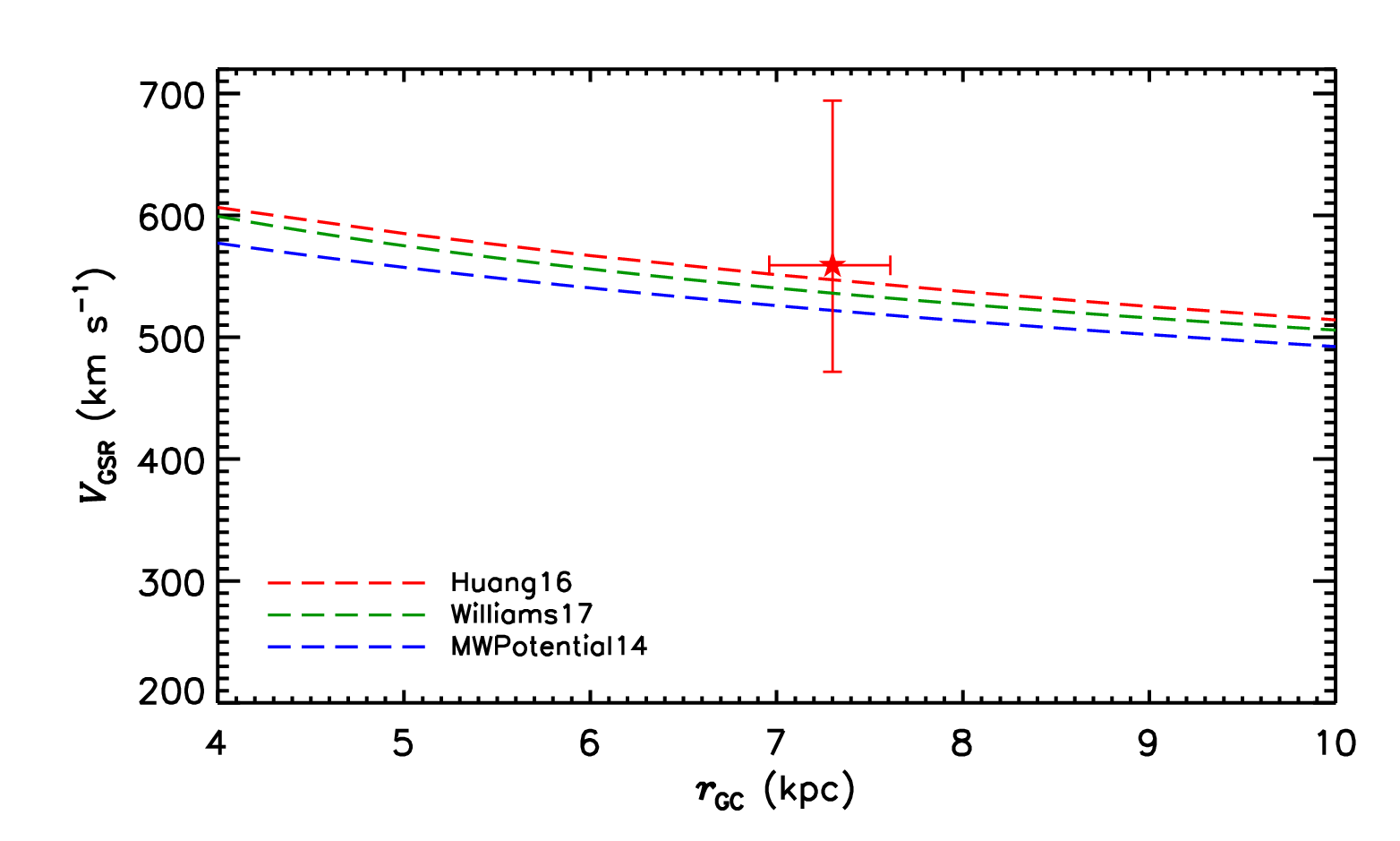}
\caption{Left panel: Average spectrum of J1443+1453 from the SUGUE survey. A synthetical spectrum (in red), of $T_{\rm eff} = 6500$\,K, log\,$g = 4.0$\,dex, [Fe/H]\,$= -2.0$\,dex and [$\alpha$/Fe]\,$=0.20$\,dex, is over-plotted for comparison.
Right panel: The Galactic rest-frame total velocity of J1443+1453 compared with the escape velocity curves, either derived directly (green dashed line; Williams et al. 2017) or predicted by the Galactic potential models (red and blue dashed lines; Bovy et al. 2015; Huang et al. 2016).}
\end{center}
\end{figure*}

\begin{figure}
\begin{center}
\includegraphics[scale=0.475,angle=0]{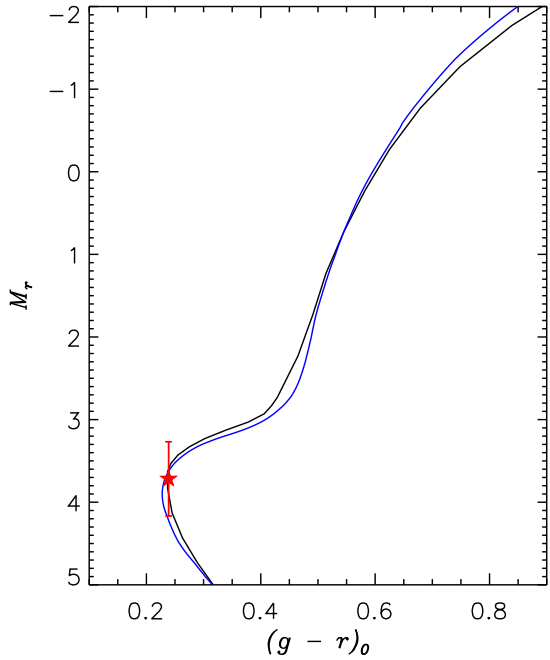}
\caption{Color-absolute magnitude diagram for J1443+1453 (red star). Black line represents an empirical isochrone of globular cluster NGC\,5466. Blue line represents a theoretical  isochrone (with [Fe/H]\,=\,$-2.0$\,dex, [$\alpha$/Fe]\,=\,$0.20$\,dex and $\tau = 14.0$\,Gyr) taken from the Dartmouth Stellar Evolution Program (Dotter et al. 2008).}
\end{center}
\end{figure}

 \begin{figure*}
\begin{center}
\includegraphics[scale=0.132,angle=0]{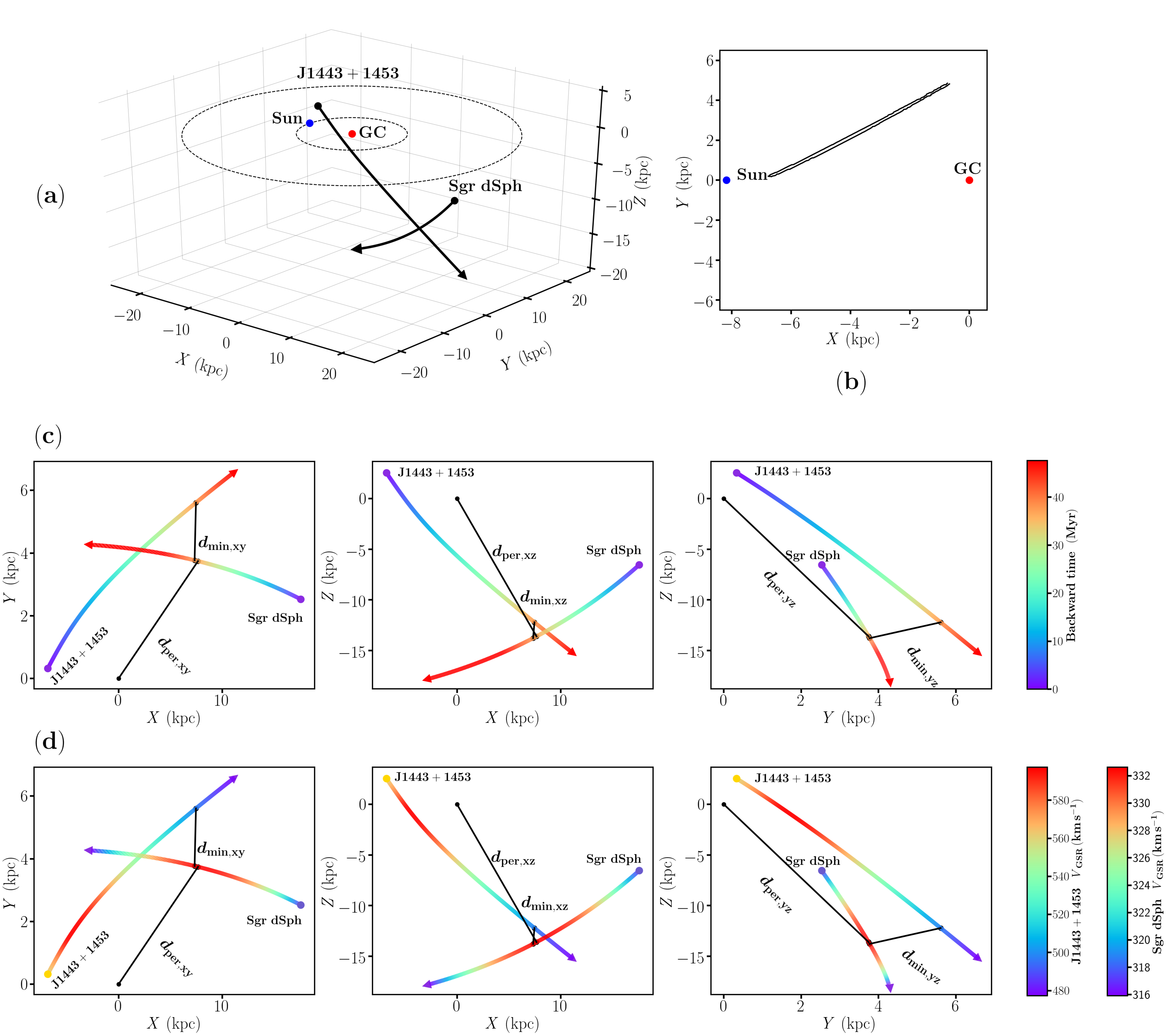}
\caption{Panel (a): Three-dimensional representation of the backward orbits of J1443+1453 and the Sgr dSph. The arrows indicate the directions of the backward orbits. The blue and red dots represent the positions of the Sun and the Galactic centre. The Solar circle ($R = 8.178$\,kpc) and the edge of the MW disk ($R = 25$\,kpc) are marked by the inner and outer dotted grey lines, respectively. 
Panel (b): Density of simulated trajectories (black contour\,=\,90 per cent confidence region) where J1443+1453 crosses the Galactic plane, in Cartesian coordinates. The blue and red dots again represent the positions of the Sun and the Galactic centre.
Panel (c): 3D orbits of J1443+1453 and the Sgr dSph, projected in $X-Y$, $Y-Z$ and $X-Z$ planes and color coded by the backward time as indicated by the right colorbar. In each sub-panel, two black solid lines are drawn, one linking the Galactic centre and the pericentre of Sgr dSph, and another linking J1443+1453 and the Sgr dSph core when the backward integrated orbits of the two intersect at the closest point.
Panel (d): Same as Panel (c) but color coded by the total velocities in the Galactic rest-frame of J1443+1453 and the Sgr dSph as indicated by the right colorbars.}
\end{center}
\end{figure*}

 \begin{figure}
\begin{center}
\includegraphics[scale=0.35,angle=0]{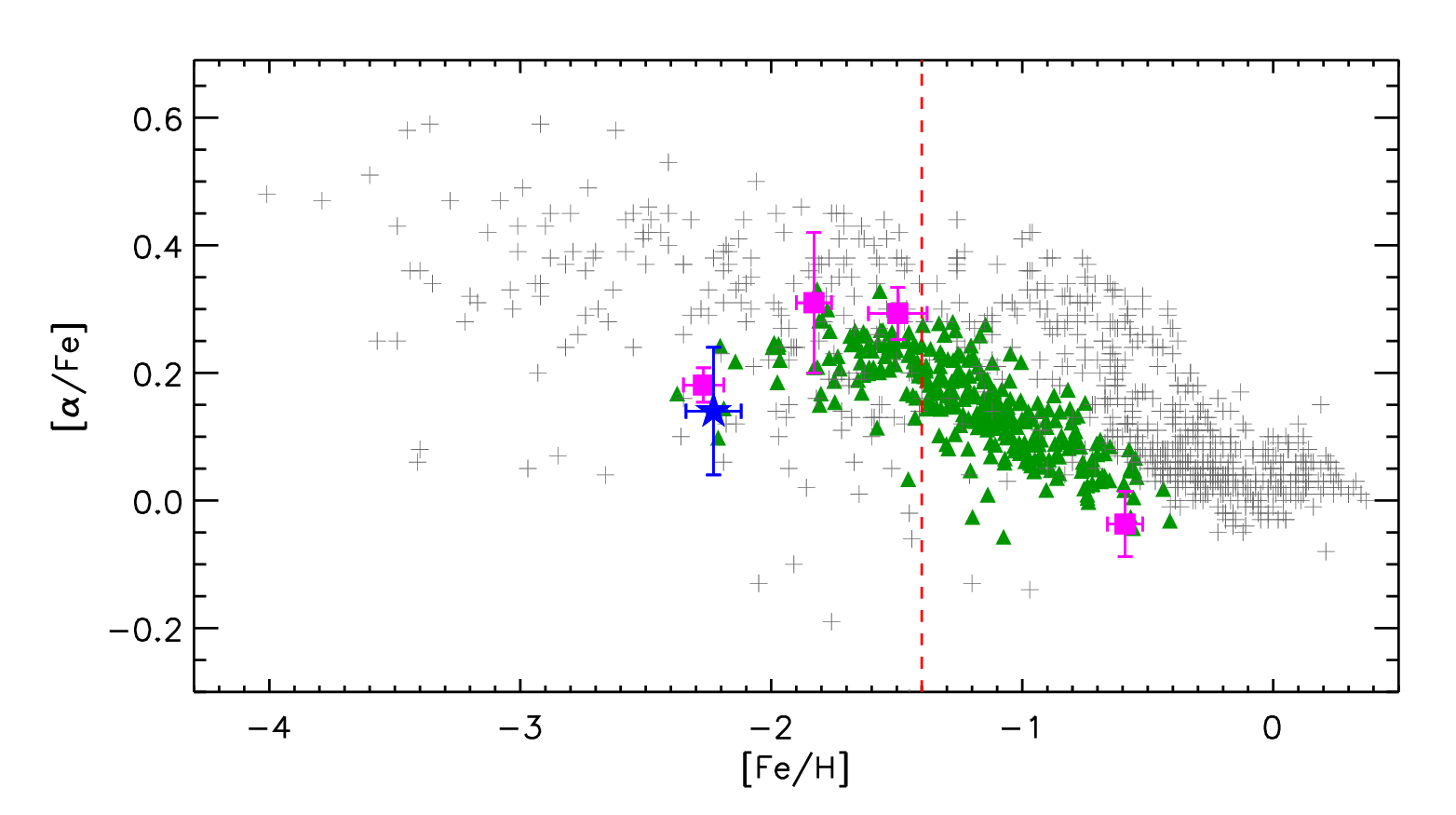}
\caption{[$\alpha$/Fe] abundance ratios as a function of [Fe/H] for J1443+1453 (blue star), Sgr dSph associated globular clusters (magenta squares), Sgr stream member stars (green triangles) and field stars of the Milk Way (grey pluses). The red dashed line marks the rough position of the $\alpha$-element `knee' of the Sgr stream. The four globular clusters associated with the Sgr dSph are  M\,54, Terzan\,7, Terzan\,8 and Arp\,2, respectively.
Their elemental abundance ratios are all taken from the results based on high resolution spectroscopy (Sbordone et al. 2007; Carretta et al. 2010, 2014; Mottini et al. 2008).}
\end{center}
\end{figure}

In addition to the  Galactic centre origin, several alternative mechanisms capable of ejecting HVSs  have been proposed, including ejected companions of Type Ia supernovae (SNe Ia; Wang \& Han 2009) and the result of dynamical interaction between multiple stars (e.g., Gvaramadze et al. 2009). 
The two mechanisms are supported by the discovery of HVSs US708 (Geier et al. 2015) and HD\,271791 (Heber et al. 2008), respectively.
Moreover, HVSs could also originate from the MW's satellite galaxy, either by tidally stripping (Abadi, Navarro \& Steinmetz 2009) or gravitational slingshot effect (e.g., Boubert \& Evans; Garc{\'\i}a-Bellido 2017; Montanari et al. 2019), assuming the satellite galaxy hosts a central massive/intermediate-mass black hole or a central massive primordial black hole (PBH).
Currently, the HVS HE\,0437-5439 is suggested to be ejected from the Large Magellanic Cloud (LMC) through Hills mechanism, requiring a massive black hole with mass of at least $4 \times 10^3$-$10^4$\,$M_{\odot}$ (Erkal et al. 2019).
More recently, Montanari et al. (2019) conducted a systematic search of candidate HVSs ejected due to close encounters between stars and PBH in the dense environments of dwarf spheroidals, in the {\it Gaia}\,DR2 HVS sample. 
However, no confident candidates are found.

In this letter, we report the discovery of a candidate HVS, J1443+1453, probably originated from the Sagittarius Dwarf Spheroidal galaxy (Sgr dSph).
In Section\,2, we present the properties of J1443+1453. 
In Section\,3, the possible origins are explored. 
Finally, we summarize in Section\,4.

\section{J1443+1453 properties}
%this star
\subsection{Spectroscopy}
J1443+1453 was spectroscopically observed twice, with a short time interval of one week between, by the Sloan Extension for Galactic Understanding and Exploration survey (SEGUE; Yanny et al. 2009).
Heliocentric radial velocities (HRVs) and stellar atmospheric parameters (effective temperature $T_{\rm eff}$, surface gravity log\,$g$ and metallicity [Fe/H]) derived with the SEGUE Stellar Parameter Pipeline (SSPP; Lee et al. 2008a) from the two spectra are almost identical and their final weighted means (by measurement errors) are presented in Table\,1 (the uncertainties listed here are typical values yielded by SSPP; Lee et al. 2008b; Lee et al. 2011).
It was a F-type ($T_{\rm eff} = 6405$\,K) star with a very low metallicity ([Fe/H]\,$= -2.23$).
The measured HRV $194.25 \pm1.97$\,km\,s$^{-1}$ corresponds to a Galactic rest-frame radial velocity of $228.22\pm1.97$\,km\,s$^{-1}$.
The average SEGUE spectrum (weighted mean by flux uncertainties) of J1443+1453  is shown in the left panel of Figure\,1.
For comparison, a synthetical spectrum taken from the G{\"o}ttingen spectral library (Husser et al. 2013) of atmospheric parameters similar to those of J1443+1453 as derived from the SEGUE spectra is overplotted, showing the robustness of the derived atmospheric parameters.

\subsection{Astrometry}
Apart from the spectroscopic observations, accurate astrometric measurements are available for J1443+1453 from the {\it Gaia} DR2 (Gaia Collaboration et al. 2018; Lindegren et al. 2018).
We estimate the distance to J1443+1453 from the {\it Gaia} parallax with a Bayesian approach,
\begin{equation}
P (d|\varpi, \sigma_{\varpi}) \propto P (\varpi|d, \sigma_{\varpi}) \times d^2P(r_{\rm GC}).
\end{equation}
Here, the parallax likelihood is given by,
\begin{equation}
P (\varpi|d, \sigma_{\varpi}) = \frac{1}{\sqrt{2\pi}\sigma_{\varpi}}\exp\frac{-(\varpi - \varpi_{\rm ZP} -\frac{1}{d})^2}{2\sigma^2_{\varpi}},
\end{equation}
where $\varpi_{\rm ZP}$ is the zero-point of the {\it Gaia} DR2 parallax and we adopt a value of $-0.048$\,mas  as determined statistically recently (Sch{\"o}nrich, McMillan \& Eyer 2019).
We set the density prior $P (r_{\rm GC}) \propto r^{-3.39}_{\rm GC}$ (McMillan et al. 2018), given the halo-star nature of J1443+1453.
A distance of $2.90^{+0.72}_{-0.48}$\,kpc is deduced for J1443+1453. 

Combing the distance derived above and the celestial coordinates, proper motions from the {\it Gaia} DR2 and HRV from the SEGUE, the 3D position and velocity of J1443+1453 are further derived.
To do so, we adopt Galactocentric distance of the Sun, $R_{0} = 8.178$\,kpc, as measured by the Gravity Collaboration (Gravity Collaboration et al. 2019), and vertical displacement of the Sun from the disc mid-plane, $Z_{\odot} = 25$\,pc (Bland-Hawthorn \& Gerhard 2016).
The Solar motions with respect to the local standard of rest adopted here are ($U_{\odot}$, $V_{\odot}$, $W_{\odot}$)\,$=$\,($7.01$, $10.13$, $4.95$)\,km\,s$^{-1}$ (Huang et al. 2015).
For the circular speed at the Solar position $V_{\rm c} (R_0)$, we use a value of $225$\,km\,s$^{-1}$, in concordance with the recent measurements (Bovy et al. 2012; Huang et al. 2016; Bland-Hawthorn \& Gerhard 2016).
The errors of the resultant 3D position and velocity are all derived by Monte Carlo (MC) simulations, by sampling the observational uncertainties of HRV from the SEGUE, proper motions from the {\it Gaia} DR2 and the distance posterior probability distribution function (PDF) derived above.
The 3D position and velocity of J1443+1453 thus deduced in a right-handed Galactocentric Cartesian coordinate system are  respectively ($X$, $Y$, $Z$)\,=\,($-6.82^{+0.32}_{-0.22}$, $0.32^{+0.08}_{-0.05}$, $2.57^{+0.60}_{-0.42}$)\,kpc and ($V_{X}$, $V_{Y}$, $V_{Z}$)\,=\,($-292.29^{+60.34}_{-98.74}$, $-183.39^{+74.27}_{-111.42}$, $439.80^{+67.45}_{-41.21}$)\,km\,s$^{-1}$.
The upper and lower uncertainties correspond to the 16 and 84 per cent percentiles of the final PDF yielded by the MC simulations.
Here $X$ passes through the Sun and points towards the Galactic centre, $Y$ is in the direction of the Galactic rotation and $Z$ points towards the north Galactic pole. 

The above results show that J1443+1453 has a total velocity in the Galactic rest-frame $V_{\rm GSR}$ of $559.01^{+135.07}_{-87.40}$\,km\,s$^{-1}$, and locates at a Galactocentric distance $r_{\rm GC}$ of $7.3^{+0.31}_{-0.34}$\,kpc.
The velocity is larger than the escape speed measured directly (Williams et al. 2017) or predicted by the potential models of the MW (Bovy et al. 2015; Huang et al. 2016) at $r_{\rm GC} = 7.3$\,kpc, placing J1443+1453 a  HVS candidate\footnote{Considering the measured total velocity uncertainty, J1443+1453 also has 43\% possibility to be a bound star, if adopting the escape velocity curve from William et al. (2017).} (see the right panel of Figure\,1).

\begin{table*}
\caption{Top three systems sorted by the value of $d_{\rm min}/r_h$ (from small to large) in the  backward orbital analysis of J1443+1453.}
\centering
\begin{threeparttable}
\begin{tabular}{ccccc}
\hline
Name&Closest distance ($d_{\rm min}$)&Backward time&Half-light radius ($r_h$)&$d_{\rm min}/r_h$\\
&(kpc)&(Myr)&(kpc)&\\
\hline
Sgr dSph&2.42&37.8&2.59\tnote{a}&0.93\\
Tucana II&16.45&291.2&0.17\tnote{b}&96.76\\
Ursa Major II&34.08&3.4&0.15\tnote{a}&228.73\\
\hline
\end{tabular}
\begin{tablenotes}
\item[a] McConnachie 2012
\item[b] Koposov et al. 2015
\end{tablenotes}
\end{threeparttable}
\end{table*}

\subsection{Age and mass}  
The age and mass of J1443+1453 are derived again by a Bayesian approach, based on the observational constraints and stellar isochrones taken from the  Dartmouth Stellar Evolution Program (DSEP; Dotter et al. 2008).
The observational constraints include the metallicity derived from the SEGUE spectra, the photometric colors from the SDSS imaging survey (after corrected for the dust reddening from Schlegel, Finkbeiner \& Davis 1998) and the $r$-band absolute magnitude inferred from the distance deduced above and the SDSS photometry (again corrected for the dust reddening).
In the estimation, the PDF of the parameters to be determined is assumed to have the form, 
\begin{equation}
f(\tau, M) = NP(\tau, M)L(\tau, M),
\end{equation}
where $N$ is a normalization factor to ensure $\iint f(\tau, M)d\tau dM = 1$.
$P (\tau, M)$ represents the priors on age and mass, and we have adopted a uniform prior for the age and a Salpeter initial mass function (Salpeter 1955) for the mass.
The likelihood function $L$ is given by,
\begin{equation}
L = \prod_{i=1}^{n}\frac{1}{\sqrt{2\pi}\sigma_i}\times\exp(-\chi^2/2),
\end{equation}
where 
\begin{equation}
\chi^2 = \sum_i^n (\frac{O_i - P_i (\tau, M)}{\sigma_i})^2.
\end{equation}
Here $O$ denotes the aforementioned observational constraints and $P$ denotes the values given by the isochrone for given $\tau$ and $M$.
A constant value 0.20\,dex of $\alpha$-element to iron abundance ratio [$\alpha$/Fe], close to that measured for J1443+1453, is adopted for all the isochrones.
The above analysis yields a mass of $0.75^{+0.03}_{-0.01}$\,$M_\odot$ and an almost cosmic age of $14.0^{+0.5}_{-2.0}$\,Gyr for J1443+1453, where the values and the uncertainties correspond to the 
50 per cent, and the 16 and 84 per cent percentiles of the resultant posterior PDF yielded by the Bayesian approach.
To show the robustness of the results, we compare the position of J1443+1453 with isochrone of [Fe/H]\,=\,$-2.0$\,dex, [$\alpha$/Fe]\,=\,$0.20$\,dex and $\tau = 14.0$\,Gyr in the $M_{r}$ versus $(g-r)_0$ plane in Figure\,3.
An empirical isochrone (An et al. 2008) of globular cluster NGC\,5466 ($d = 16.0$\,kpc, [Fe/H]\,=\,$-1.98$\,dex and $\tau = 13.0 \pm 0.75$\,Gyr; Harris 2010; Dotter 2010) is also overplotted.   
The plot shows that J1443+1453 falls closely near the main-sequence turn-off (MSTO) region, for both the theoretical and the empirical isochrones.

 \section{The possible origins of J1443+1453}

%the origin of J1443+1453
\subsection{Backward orbit analysis}
 To constrain the possible ejection location of J1443+1453, we perform a backward orbital integration in a model Galactic potential with the package {\tt Gala} (Price-Whelan 2017).
 We use the classical Galactic potential model {\tt MWPotential2014} (Bovy et al. 2015) consisting of three components, a bulge, a disk and a dark matter halo.
 The orbit is integrated backwards in time step of 0.1\,Myr.
A representation of the integrated orbit in 3D space is shown in Figure\,3a.
 
 %1: GC origin
We first  check if this HVS was ejected  by the central SMBH of the MW.
We find that J1443+1453 intersects the Galactic plane ($Z=0$\,kpc) $5.6^{+0.4}_{-0.6}$\,Myr ago at location ($X$, $Y$)\,=\,($-5.07^{+1.04}_{-0.73}$, $1.34^{+0.80}_{-0.51}$)\,kpc (see Figure\,3b).
The intersection is thus too far from the Galactic centre to make it an HVS created by the Hills Mechanism.

%GC or dwarf galaxies origin:
To further explore the possible origin of J1443+1453, we integrate the backward trajectories of 150 Galactic globular clusters and 39 dwarf galaxies with full information of  phase-space positions and motions (Vasiliev 2019; Fritz et al. 2018).
Here the orbit is integrated up to 5\,Gyr back in time with a step of 0.1\,Myr.
We note that the gravitational influence of the globular cluster or dwarf galaxy is ignored in our orbital analysis.
The possible link of the trajectory of J1443+1453 and those of the globular clusters and dwarf galaxies are investigated by sorting the closest orbital distance to half-light/mass radius ratio (see Table\,2 for the top 3 systems).
Excitingly, our backward orbital analysis shows that the orbit of J1443+1453 intersects with that of Sgr dSph within its half-light radius $r_{\rm h} = 2.59$\,kpc (McConnachie 2012) $37.8^{+4.6}_{-6.0}$\,Myr ago.
The closest encounter occurred at an impact distance $2.42^{+1.80}_{-0.77}$\,kpc from the core of Sgr dSph.
Moreover, the encounter occurred when the Sgr dSph was at position ($X$, $Y$, $Z$)\,=\,($7.37^{+1.74}_{-1.39}$, $3.77^{+0.79}_{-0.72}$, $-13.76^{+1.54}_{-1.44}$)\,kpc, very close to the pericentre  ($X$, $Y$, $Z$)\,=\,($7.70^{+1.63}_{-1.63}$, $3.74^{+0.61}_{-0.65}$, $-13.58^{+0.74}_{-0.60}$)\,kpc (see Figure\,3c and 3d) during its latest passage of the MW.
At the closest approach, J1443+1453 had a velocity of $689.69^{+103.53}_{-64.72}$\,km\,s$^{-1}$ relative to the Sgr dSph (see Figure\,3d).

In the current backward orbital analysis, J1443+1453 is 0.97\,kpc away from the center of Sgr dSph, during the closest encounter, at $3\sigma$ confidence.
Even considering the gravitational influence of Sgr dSph (assuming Sgr dSph as a $10^9 M_{\odot}$ Keplerian flyby; Erkal \& Belokurov 2015), the effect on estimating the closest encounter distance is smaller than 0.2\,kpc.

We repeat the whole orbit integration analysis by adopting alternative assumptions to test the robustness of our results.
First, we change the Solar motions to ($U_{\odot}$, $v_{\odot}$, $W_{\odot}$)\,$=$\,($11.10$, $250.00$, $7.24$)\,km\,s$^{-1}$ (Sch{\"o}nrich et al. 2010, 2012) and keep other parameters unchanged. 
The orbit of J1443+1453 intersects with that of the Sgr dSph within its half-light radius $37.8^{+4.6}_{-5.9}$\,Myr ago.
The closest encounter is $2.52^{+1.89}_{-0.72}$\,kpc away from the Sgr dSph core and just around its pericentre. 
As another test, we change to the default  Galactic  potential model of {\tt Gala} that consists of  four components (nucleus, bulge, disk and dark matter halo) and again keep other parameters unchanged.
Again, J1443+1453 had a close meet with the Sgr dSph at a distance of $2.40^{+1.83}_{-0.69}$\,kpc from the core of the latter, $37.2^{+4.4}_{-5.6}$\,Myr ago.
The tests show that the impact of alternative assumptions on the above orbital analysis is quite limited.

The upper and lower uncertainties of those reported parameters from our orbital analysis come from 16 and 84 per cent percentiles of the PDF yielded by 10,000 MC trajectory calculations, assuming that the measurement errors are normally distributed except the distance (for which the posterior PDF derived above is used directly).

\subsection{Chemical abudnaces}
We show J1443+1453 in the [Fe/H] versus [$\alpha$/Fe] plane in Figure\,3.
The location is consistent with the distribution of the Sgr stream member stars and differs significantly from that of the Galactic field stars (Venn et al. 2004).
Member stars of the Sgr stream are taken from an analysis based on the LAMOST K giants (Yang et al. 2019).
Their elemental abundance ratios have been obtained  by applying a data-driven Payne approach to the LAMOST low-resolution spectra (Xiang et al. 2019) of signal-to-noise rations greater than 25 in $g$-band. 
More interestingly,  the chemical composition of J1443+1453  is very close to globular cluster Terzan\,7, known to be associated with the Sgr dSph.
The result suggests that J1443+1453 could come from a halo star of the Sgr dSph, in line with the above conclusion from the orbital analysis.

\subsection{Origin of J1443+1453}
The above orbital and chemical analysis strongly suggested J1443+1453 is originated from the Sgr dSph.
Here we discuss the possible ejection mechanisms of J1443+1453 from the Sgr dSph.

First, HVSs could be ejected from disrupting dwarf galaxies by tidally stripping during the pericentric passage of the latter trough the MW (Abadi, Navarro \& Steinmetz 2009).
The backward orbit analysis shows that the orbit of J1443+1453 intersects closely with that of the Sgr dSph $37.8^{+4.6}_{-6.0}$ Myr ago, when the latter has its latest pericentric passage through the MW.
This finding is in excellent agreement with the theoretical prediction by Abadi, Navarro \& Steinmetz (2009), and strongly suggest that J1443+1453 is an HVS probably stripped from the
tidally disrupting Sgr dSph. 
As a next step, further numerical simulation is required to investigate the possibility of ejecting J1443+1453 like HVS from the Sgr dSph by tidally stripping.

On the other hand, J1443+1453 can also be ejected from the Sgr dSph by the gravitational slingshot effect (e.g., Hills 1988; Garc{\'\i}a-Bellido 2017; Montanari et al. 2019),  if the later is confirmed to host a (central)  massive/intermediate-mass black hole or a (central)  massive PBH.
By assuming the Hills mechanism  (Hills1988;  Bromley et al. 2006), the mass of the black hole is at least $1.56^{+2.05}_{-0.70} \times 10^3$\,$M_{\odot}$ to eject a HVS like J1443+1453 from the Sgr dSph with an ejection velocity of $689.69^{+103.53}_{-64.72}$\,km\,s$^{-1}$.

\section{Summary}
In this letter, we present an  intriguing HVS (J1443+1453) candidate with a total velocity of  $559.01^{+135.07}_{-87.40}$\,km\,s$^{-1}$ in the Galactic rest-frame and a heliocentric distance of $2.90^{+0.72}_{-0.48}$\,kpc.
By backward orbit analysis, this HVS is found to have a close encounter with the Sgr dSph $37.8^{+4.6}_{-6.0}$\,Myr ago, during the latest Galactic pericentric passage of the latter.
The chemical properties of  J1443+1453 also show it could be a member star of  the Sgr dSph.
These results strongly suggest that J1443+1453 is probably a halo HVS stripped from the tidally disrupting Sgr dSph during its latest pericentric passage, exactly in line with the theoretical
predictions (Abadi, Navarro \& Steinmetz 2009). On the other hand, we can not rule out the possibility that this HVS is ejected from the Sgr dSph by the gravitational slingshot effect (e.g., Hills 1988; Garc{\'\i}a-Bellido 2017), by assuming a (central) massive/intermediate-mass black hole or a (central)  massive PBH in the Sgr dSph.

We are excited by the prospect of finding more dwarf galaxy originated HVSs with the ongoing and forthcoming large-scale surveys. 
Further identifications of such objects will provide not only vital constraints on the nature and ejection mechanisms of HVSs, but also new insights for understanding galaxy formation and evolution.

\section*{ \small Acknowledgements} 
We would like to thank the referee for his/her helpful comments.
This work is supported by National Natural Science Foundation of China grants 11903027, 11973001, 11833006, 11811530289 and U1731108, and  National Key R\&D Program of China No. 2019YFA0405500. 
Y.H. is supported by the Yunnan University grant C176220100007. 
We used data from the European Space Agency mission Gaia (\url{http://www.cosmos.esa.int/gaia}), processed by the Gaia Data Processing and Analysis Consortium (DPAC; see \url{http://www.cosmos.esa.int/web/gaia/dpac/consortium}). 
We also used the data from the SDSS survey.

\end{document}